\documentclass[journal]{IEEEtran}
\usepackage{graphicx}
\usepackage[font={small}]{caption}	
\usepackage[font=small,labelfont={small}]{caption}
\usepackage{amsmath}

\begin{document}
\title{Non-Invasive Induction Link Model for Implantable Biomedical Microsystems: \emph{Pacemaker} to Monitor Arrhythmic Patients in Body Area Networks}

\author{A. Tauqir$^{1}$, S. Akram$^{1}$, A. H. Khan$^{2}$, N. Javaid$^{2,3}$, M. Akbar$^{2}$\\\vspace{0.4cm}
$^{1}$Institute of Space Technology, Islamabad, Pakistan.\\
$^{2}$Dept. of Electrical Engineering, COMSATS Institute of IT, Islamabad, Pakistan.\\
$^{3}$CAST, COMSATS Institute of IT, Islamabad, Pakistan.}

\maketitle

\begin{abstract}
In this paper, a non-invasive inductive link model for an Implantable Biomedical Microsystems (IBMs) such as, a pacemaker to monitor Arrhythmic Patients (APs) in Body Area Networks (BANs) is proposed. The model acts as a driving source to keep the batteries charged, inside a device called, pacemaker. The device monitors any drift from natural human heart beats, a condition of arrythmia and also in turn, produces electrical pulses that create forced rhythms that, matches with the original normal heart rhythms. It constantly sends a medical report to the health center to keep the medical personnel aware of the patient's conditions and let them handle any critical condition, before it actually happens. Two equivalent models are compared by carrying the simulations, based on the parameters of voltage gain and link efficiency. Results depict that the series tuned primary and parallel tuned secondary circuit achieves the best results for both the parameters, keeping in view the constraint of coupling co-efficient (k), which should be less than a value \emph{0.45} as, desirable for the safety of body tissues.
\end{abstract}

\begin{IEEEkeywords}
Heart, arrythmia, pacemaker, inductive link, implant, voltage gain, link efficiency.
\end{IEEEkeywords}

\section{Introduction}
Since many years, Medical Science is putting great effort in developing and implanting variable devices that are placed inside or on the surface of the body. Many implants are prosthetics, i.e. intend to replace missing body parts. Other implants deliver medication, monitor body functions, or provide support to organs and tissues.

These implants aim to provide instant medical aid and take control over an abnormal function of a specific organ. Implants are either made from skin, bone and other body tissues or are made from metal, plastic, ceramic and other materials. According to the need, they are either permanently placed or temporarily placed i.e. removed, once they are no longer needed. Some require charging while, others do not. The devices that require power for their operation, deplete after certain days or weeks or years. To re-charge or replace the dead batteries of such devices, it is not a good practise to operate the human body very often, mainly, due to health and cost factors. To avoid such issues, induction mechanism is used to keep the batteries charged, keeping in mind, the certain constraints which, normally include heat radiations generating from the inductive coils that cause severe damage to the body tissues.

According to the statistics, the most widely implanted device inside the human body is a Pacemaker. The second most important and sensitive human organ after brain, is the heart. The ratio of heart related problems as compared to any other organs diseases, is significantly high among the people, world-wide.

Heart is one of the most important organs in a human body which, is a muscular organ that pumps blood to the body. It weighs approximately between \emph{250} to \emph{300} $grams$ and \emph{300} to \emph{350} grams in females and males, respectively. It is around the size of the fist. It pumps fresh blood to the whole body and after providing oxygen to every cell; it carries back the de-oxygenated blood, back to the heart. This process is called the pumping of a heart and a normal human heart beats, \emph{72} times a minute. It normally pumps almost \emph{2000} $gallons$ of blood, a day. The heart consists of four chambers:

\begin{itemize}
  \item \emph{Right atrium:} receives blood from the veins and pumps it to the right ventricle.
  \item \emph{Right ventricle:} then, pumps the blood to the lungs where, it is loaded with oxygen.
  \item \emph{Left atrium:} receives the oxygenated blood from the lungs and then, pumps it to the left ventricle.
  \item \emph{Left ventricle:} pumps oxygen-rich blood to the rest of the body and hence, every cell gets proper ratio of oxygen and nutrition.
\end{itemize}

The heart possesses an intrinsic electrical system that, controls the rate and rhythm of the heartbeat. An electrical signal spreads from top of the heart to the bottom, upon each heart beat. As the signal travels, it causes the heart to contract and pump the blood. Each signal normally begins in a group of cells, called the sinus node. Due to this signal, the heart's upper two chambers, the atria contract which, pumps blood into the heart's two lower chambers, the ventricles. The ventricles then contract and pump the blood to the rest of the body. This combined contraction of the atria and ventricles is a heartbeat.

There are different heart diseases. Some of them are mentioned below:
\begin{itemize}
  \item \emph{Arrhythmia:} refers to an abnormal heart rhythm, due to changes in the conduction of electrical impulses through the heart. Some arrhythmias are benign, however, others are life-threatening. The heart may not be able to pump enough blood to the body and also, lack of blood flow can, damage the brain, heart, and other organs.
  \item \emph{Cardiomyopathy:} refers to a disease of heart muscle in which, the heart is abnormally enlarged, thickened, or stiffened. As a result, the heart's ability to pump blood to the whole body gets weakened.
  \item \emph{Pulmonary embolism:} refers to a blood clot that, travels through the heart to the lungs.
  \item \emph{Stable angina pectoris:} refers to the chest pain or discomfort with exertion that, creates blockages and prevents the heart from receiving the extra oxygen, needed for hard activities.
\end{itemize}

One of the aforementioned issues namely, arrhythmia, is handled by implanting a device called, pacemaker inside the human body worldwide. In this paper, the aim is to provide an induction model to recharge the batteries of the sensors inside a pacemaker.

The remainder of the paper is organized as follows. Section II gives an overview about the working of pacemaker. Related work is mentioned in section III. In section IV, motivation for proposing the induction model is mentioned. Parameters for induction link and mathematical modeling are explained in section V. Section VI, discusses and compares the plots between the models. Section VII concludes the paper. Future work is mentioned in Section VIII.

\section{Pacemaker}
Faulty electrical signaling in the heart causes arrhythmias. Pacemakers use low-energy electrical pulses to overcome this faulty electrical signaling. They create forced rhythms according to natural human heart beats, to let the heart to function in a normal manner. They are also used for the following purposes as:
\begin{itemize}
  \item Speed up a slow heart rhythm and help control, an abnormal or fast heart rhythm.
  \item Coordinate electrical signaling between the upper and lower chambers of the heart.
  \item May act as cardiac re-synchronization therapy (CRT) devices, that coordinate electrical signaling between the ventricles, letting them to treat heart failure.
  \item Monitor and record the heart's electrical activity, blood temperature, breathing rate, and other factors like fatigue level in human body.
  \item Adjust heart rate according to the changes in the activity of the human body, thereby, adaptive in nature.
\end{itemize}

The following subsection briefly discuss about the working and deployment of a pacemaker.

\subsection{Working of Pacemaker}
A pacemaker circuitry consists of a small battery, a generator and wires attached to the sensor to be inserted into the patient’s heart. It senses a heartbeat and sends the signals to the computer generator, by the help of wires. If the heartbeat is not normal, it generates small electrical signals to regulate the heartbeat. Internal Circuitry of pacemaker is shown in Fig. 1.
%figure-1
\begin{figure}[htbp]
  \centering
  \includegraphics[scale=0.5]{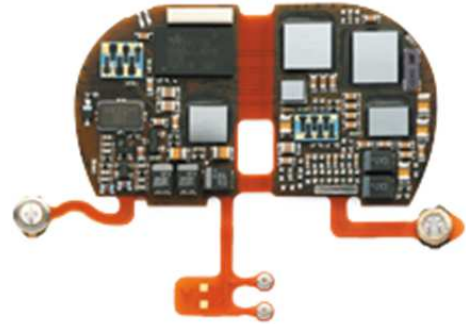}\\
  \caption{Internal view of pacemaker}\label{1}
\end{figure}

Pacemakers are either deployed, temporarily or permanently. Temporary pacemakers are used to treat short-term heart problems such as, a slow heartbeat that is caused by a heart attack, heart surgery, or an overdose of medicine. Permanent pacemakers are used to control long-term heart rhythm problems. Fig. 2 below shows that, which part of the heart is attached with the probes of the pacemaker.
%figure-2
\begin{figure}[htbp]
  \centering
  \includegraphics[scale=0.35]{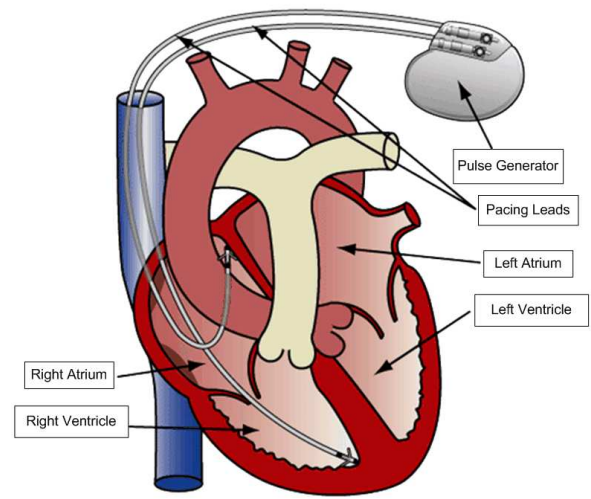}\\
  \caption{Placement of pacemaker's probes inside a human heart}\label{2}
\end{figure}

\section{Related Work}
In \cite{ali2009mathematical}, a generalized model is proposed that covers all possible voltage driven models of an inductive link for medical implants. Results are analyzed by using the parameters of voltage gain and link efficiency.

In \cite{hmida2007design}, an inductive power system is proposed, that combines power transfer with data transmission for implantable biomicrosystems. The implant receives power from an external transmitter through an inductive link, between an external power transmission coil and the implanted receiving coil.

In \cite{ali2009inductive}, a transcutaneous link is designed for medical implants, using inductively coupled coils. It also describes the design of an indigenously developed transcutaneous link from commercial off-the-shelf components to demonstrate the design process.

In \cite{mehrnoush2010power}, one of the most suitable technologies for self-powered BANs in terms of magneto-inductive nodes and channels is proposed. Authors also investigate the impacts of magnetically coupled transceiver antenna coil features, on the received signal power and the communication link capacity.

In \cite{chaoui2005electrical} and \cite{zierhofer1990high}, authors propose an innovative modeling method for the mutual inductance of two magnetically coupled coils in an inductive link. It focuses on optimizing the voltage gain of the link, comprising the power amplifier plus the inductive link.

In \cite{burny2000concept}, a medical relevance of the monitoring of deformation of implants is presented, as a powerful tool to evaluate nursing and rehabilitation exercises, for tracing dangerous overloads and anticipating implant failure and also to observe the healing process.

Tahir \emph{et al.} in \cite{tahir2013adaptive}, present a technique named as, EAST. The technique uses an open-loop control to compensate for the changes in link quality, according to the variations in temperature.

An analytical model for the optimal frequency of a coil in terms of the design parameters is presented in \cite{yang2007inductor}. By varying the design parameters, the optimal frequency gets close to the operating frequency, thereby, boosting the efﬁciency of the inductive link.

Sawan \emph{et al.} in \cite{sawan2009multicoils}, propose an integrated receiver micro-coil for inductively-coupled links. The link is used to transfer energy to implantable devices within short distances, in an efficient manner.

In \cite{guo2011inductive}, an inductive coupling is considered for efficiently transmitting power in wireless networks to implantable devices. Authors also focus on optimizing the rectangular coils, used in inductive links for more general applications.

\section{Motivation}
The heart pumps blood to all parts of the body. This pumping happens with special rhythms called, natural heart beats. When a heart beat gets slower, faster or abnormal than usual, the entire body significantly suffers, thereby, leads to a deficiency of oxygen in the cells. Such disease is called, arrhythmia. A device called pacemaker is in use for decades, especially to control this disease. The probes from the pacemaker gets connected to the inside of the muscles of the heart and generates electrical pulses. Sensed information of different events occurring inside the heart, are immediately transmitted to the doctor. All this processing and transmitting of data, create a strain on the battery of a pacemaker to consume huge amount of power. Hence, the sensor depletes and becomes unable to further carry any informational data. This situation is life threatening for the patients. Keeping this in mind, many researches are conducted to propose induction links which, focus on minimizing the problem of inefficient transfer of induction from primary side to secondary side and ultimately, helps in recharging the sensors.

In this paper, an induction technique is presented, to recharge the sensors battery, implanted inside a pacemaker without letting the patient to undergo any frequent surgery. it focus on enhancing voltage gain and link efficiency.

\section{Mathematical Model}
An inductive link consists of a primary circuit that is powered by a voltage source which then, generates magnetic flux in order to induce power at secondary side, implanted inside human body. The skin acts as an interface or a barrier between the two circuits. A parameter known as, coupling co-efficient (k), is a degree of coupling between the two circuits. It is a very important factor in enhancing the efficiency of the link. For WBANs, k should be less than a value of \emph{0.45}, in order to avoid damage to the body tissues. Two parameters namely, voltage gain and link efficiency are used for the validation of an inductive link. Voltage gain refers to a ratio that, indicates an increase in the voltage at the output side in relative to the voltage applied at primary side. Link efficiency refers to the ability of transferring power from primary side to secondary side in an efficient manner. Both the parameters are highly dependent upon the factor, k.

The following subsection discusses the equations for the two equivalent circuits, used for induction.

\subsection{Equivalent Circuits}
Two equivalent circuits namely, series tuned primary circuit and series tuned primary and parallel tuned secondary circuit are discussed in detail in the following sections.

\subsubsection{Series Tuned Primary Circuit}

In series tuned primary circuit, a capacitor is connected in series at primary side. As, only a small amount of voltage induces because of a low coupling factor of \emph{0.45} so, a series tuned circuit is used in order to induce sufficient amount of voltage to the secondary coil. The circuit is shown in Fig. 4.
%figure-12
\begin{figure}[htbp]
  \centering
  \includegraphics[scale=0.4]{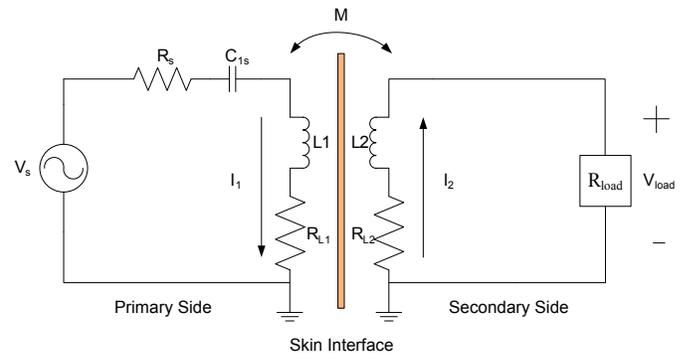}\\
  \caption{Series tuned primary circuit}
\end{figure}

$V_{s}$ is the source voltage with $R_{s}$ as the series resistance. $C_{1s}$ is the series resonating capacitor at primary side and $L1$ and $L2$ are the inductors at primary and secondary side with their series resistance of $R_{L1}$ and $R_{L2}$, respectively. $M$ is the mutual inductance and $R_{load}$ is the load resistance across which, the output voltage $V_{load}$ is measured which then, drives the implanted sensors or a device. The values of the parameters are shown below in TABLE I.
% Table-3
\begin{table}[h]
\centering
\begin{tabular}{c|c}
  \hline\hline
  \bf{Parameters} & \bf{Values}\\
  \hline
  Operating frequency & f=13.56 MHZ \\
  Primary coil & L$_{1}$ = 5.48 $\mu$H\\
  Secondary coil & L$_{2}$ = 1 $\mu$H\\
  Parasitic resistance of the transmitter coil & R$_{L1}$ $\simeq$ 2.12 $\Omega$ \\
  Parasitic resistance of the receiver coil & R$_{L2}$ $\simeq$ 1.63 $\Omega$  \\
  Load resistance & R$_{load}$ = 320 $\Omega$  \\
  \hline\hline
\end{tabular}
\caption{Inductive link parameters}
\end{table}

The equations for primary side are as follows:

By applying Kirchoff's Voltage Law, source voltage is given as:

\begin{equation}
 V_{s} = I_{1}A - j\omega M I_{2}
\end{equation}

here, $A = R_{s}+R_{L1}+j\omega L_{1} + \frac{1}{j\omega C_{1s}}$\\

%By re-arranging the terms,
%\begin{equation}
% I_{1} = \frac{V_{s} + j\omega M I_{2}}{A}
%\end{equation}

The equations for secondary side are as follows:

The output voltage, $V_{load}$ is given as:

\begin{equation}
 V_{load} = - I_{2}(j\omega L_{2}+R_{L2}) + j\omega M I_{1}
\end{equation}

The current in secondary side $I_{2}$, is given as:

\begin{equation}
 I_{2} = \frac {V_{load}}{R_{load}}
\end{equation}

By proper substitutions:

\begin{equation}
  \frac{V_{load}}{V_{s}} = \frac {j\omega M R_{load}} {AB+\omega^{2}M^{2}}
\end{equation}

where, $B = R_{load} + j\omega L_{2} + R_{L2}$ \\

The next parameter to find, is the efficiency of the induction link. It is given by the following formula as:

\begin{equation}
  \eta = \frac {P_{2}} {P_{1}} = \frac {V_{load}I_{2}}{V_{s}I_{1}} = \bigg(\frac{V_{load}}{V_{s}}\bigg) \bigg(\frac{I_{2}}{I_{1}}\bigg)
\end{equation}

By substituting the values of $\frac{V_{load}}{V_{s}}$, $I_{1}$ from equation (1) and $I_{2}$ from equation (3), link efficiency becomes:

\begin{equation}
  \eta = \bigg(\frac {j\omega M R_{load}} {AB+M^{2}\omega^{2}}\bigg) \bigg(\frac{V_{load}}{R_{load}}\bigg) \bigg(\frac{j\omega M R_{load}}{V_{load}B}\bigg)
\end{equation}

For actual consumption at the output, we take real of equation (6). Now, $\eta$ becomes:

\begin{equation}
  \eta = \frac {\omega^{2} M^{2} R_{load}} {Re[B](Re[A]Re[B] + \omega^{2}M^{2})}
\end{equation}

where, $Re[A] = R_{s} + R_{L1}$ and $Re[B] = R_{load} + R_{L2}$

\subsubsection{Series Tuned Primary and Parallel Tuned Secondary Circuit}
This circuit's primary side is the same as the primary side of series tuned circuit discussed above. However, a capacitor $C_{2p}$ is connected in parallel at secondary side as, the sensors implanted inside a human body operate under low frequencies. This parallel capacitor let the circuit to act as a low pass filter which, allows low frequencies to pass through while, blocking the higher frequencies thereby, preventing damages to body tissues. Also, this topology helps in achieving high voltage gain and better link efficiency. The circuit is shown in Fig. 5.
%figure-13
\begin{figure}[htbp]
  \centering
  \includegraphics[scale=0.4]{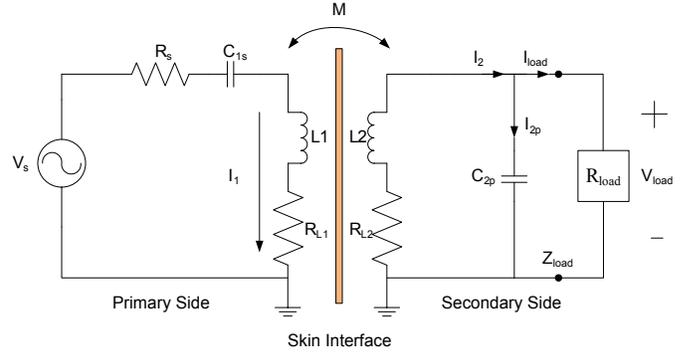}\\
  \caption{Series tuned primary and parallel tuned secondary circuit}\label{5}
\end{figure}

Equations for primary side of the circuit are the same as given in above section. However, for secondary side, the equations are discussed below.

The combined load after the secondary windings is $Z_{load}$ which, is given as:

\begin{equation}
Z_{load}=\frac{R_{load} - j\omega C_{2p}R_{load}^{2}}{1 + \omega^{2} C_{2p}^{2}R_{load}^{2}}
\end{equation}

where, $R_{2}$ is the real component and $X_{2}$ is the reactance of $Z_{load}$ given as:\\

$R_{2} = \frac {R_{load}} {1+\omega ^{2} R_{load}^{2} C_{2p}^{2}} $\\

$X_{2} = -\frac {\omega R_{load}^{2}C_{2p}} {1+\omega^{2} R_{load}^2C_{2p}^{2}}$\\

Now, solving for voltage at secondary side,

\begin{equation}
V_{load} = -I_{2}(j\omega L_{2}+R_{L2})+j\omega M I_{1}
\end{equation}

Also,

\begin{equation}
V_{load} = I_{load}R_{load} = I_{2}Z_{load}
\end{equation}

By proper substitutions, $V_{load}/V_{s}$ becomes:

\begin{equation}
 \frac{V_{load}}{V_{s}} = \frac {j\omega M Z_{load}} {A(Z_{load}+jwL_{2}+R_{L2})+\omega^{2} M^{2}}
\end{equation}

For link efficiency,

 \begin{equation}
  \eta = \frac {j^{2}\omega^{2} M^{2}Z_{load}}{C(AC+\omega^{2} M^{2})}
\end{equation}

where, $C = Z_{load}+j\omega L_{2}+R_{L2}$ \\

Taking real of equation (12), the final equation for $\eta$ becomes:

\begin{equation}
  \eta = \frac {\omega^{2} M^{2} Re[Z_{load}]} {Re[C](Re[A]Re[C] + \omega^{2}M^{2})}
\end{equation}

where, $Re[C] = Re[Z_{load}] + R_{L2}$ and $Re[Zload] = R2$

\section{Simulation Results}
Results for the above mentioned equivalent circuits are discussed in the following two subsections.
\subsection{Results for Series Tuned Primary Circuit}
From equation (4), it is clear that the voltage gain $V_{load} / V_{s}$ for series tuned primary circuit is directly proportional to $R_{load}$ as, represented in Fig. 7. As, the value of k increases from \emph{0.2} to \emph{0.8}, $V_{load} / V_{s}$ also increases, respectively. As, $R_{load}$ increases from \emph{0}\emph{$\Omega$} to \emph{400}\emph\emph{$\Omega$}, $V_{load} / V_{s}$ shows almost a linear response.
For coupling co-efficient, $k$ $=$ \emph{0.4} and at $R_{load}$ $=$ \emph{320}\emph{$\Omega$}, $V_{load} / V_{s}$ possess a value of \emph{2.5}. The aforementioned values of $k$ and $R_{load}$ \cite{ali2009mathematical}, are considered to be harmless for human tissues.

% figure-31
\begin{figure}[htbp]
  \centering
  \includegraphics[height=6.7cm,width=8.7cm]{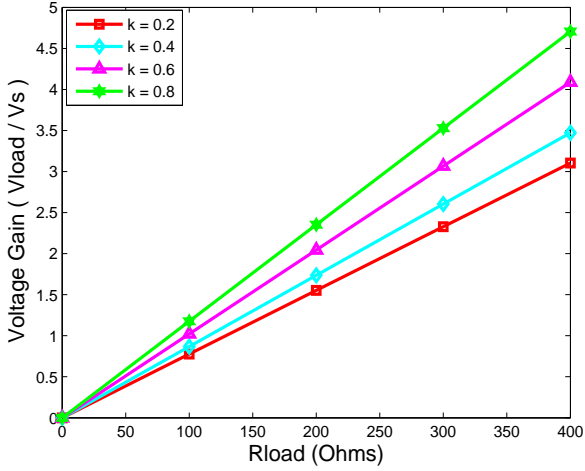}\\
  \caption{Voltage gain of series tuned primary circuit}\label{7}
\end{figure}

The result for link efficiency of series tuned primary circuit is shown in Fig. 8. It is very clear from the figure that, the value of link efficiency is highly dependent on $k$. At a value of $k$ $=$ \emph{0.4} and at $R_{load}$ $=$ \emph{320}\emph{$\Omega$}, the value of $\eta$ is \emph{0.75}, i.e. the link efficiency is about \emph{75}$\%$.
% figure-32
\begin{figure}[htbp]
  \centering
  \includegraphics[height=6.7cm,width=8.7cm]{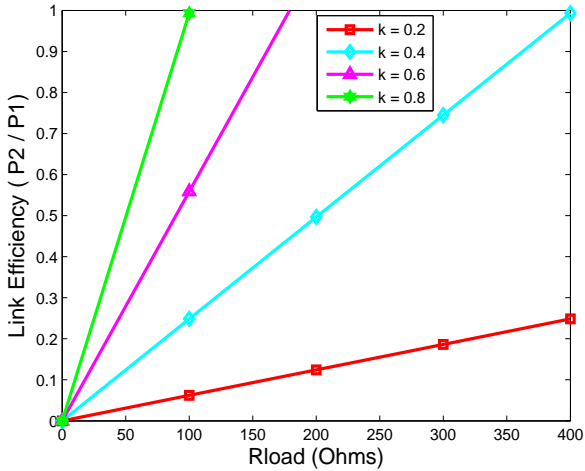}\\
  \caption{Link efficiency of series tuned primary circuit}\label{8}
\end{figure}
\subsection{Results for Series Tuned Primary and Parallel Tuned Secondary Circuit}
In Fig. 9, changes in voltage gain $V_{load} / V_{s}$, by varying $R_{load}$ for series tuned primary and parallel tuned secondary circuit is shown. It is obvious from the figure that, any change in $R_{load}$ produces no significant change in $V_{load} / V_{s}$. The only noticeable increase in $V_{load} / V_{s}$, is produced by increasing the coupling coefficient $k$. However, due to the fact that this link is used on human body to induce voltage to the implanted device, value if $k$ changes only from \emph{0} to \emph{0.45}. From $R_{load}$ $=$ \emph{0}\emph{$\Omega$} to \emph{100}\emph{$\Omega$}, the behaviour of $V_{load} / V_{s}$ is nearly constant. After which, it increases nearly linearly. For $k$ $=$ \emph{0.4} and $R_{load}$ $=$ \emph{320}\emph{$\Omega$}, the voltage gain is about \emph{3.7}. Hence, the value of $V_{load} / V_{s}$ is significantly higher when, the seconday circuit is tuned in parallel, in relative to tuning only the primary circuit in series. Also, the output voltage is nearly \emph{4} times higher than the driving input voltage.
% figure-33
\begin{figure}[htbp]
  \centering
  \includegraphics[height=6.7cm,width=8.7cm]{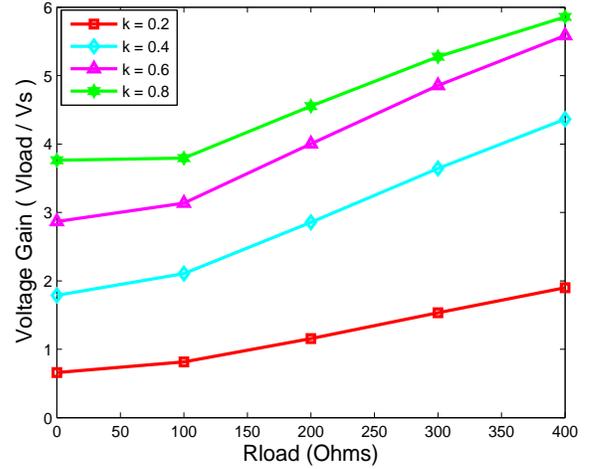}\\
  \caption{Voltage gain of series tuned primary and parallel tuned secondary circuit}\label{9}
\end{figure}

The link efficiency graph for series tuned primary and parallel tuned secondary circuit is shown in Fig. 10. This figure depicts steepness for link efficiency when, $R_{load}$ $=$ \emph{100}\emph{$\Omega$} for every value of $k$. After this value, the link efficiency of the circuit becomes constant. For $k$ $=$ \emph{0.4} and $R_{load}$ $=$ \emph{320}\emph{$\Omega$}, the link efficiency is about \emph{0.9}, i.e. \emph{90}$\%$. So, in comparison to the series tuned primary circuit, the link efficiency of the other circuit increases by \emph{15}$\%$. Also, this shows that, \emph{90}$\%$ of the input power is efficiently transferred to the secondary side.
% figure-34
\begin{figure}[htbp]
  \centering
  \includegraphics[height=6.7cm,width=8.7cm]{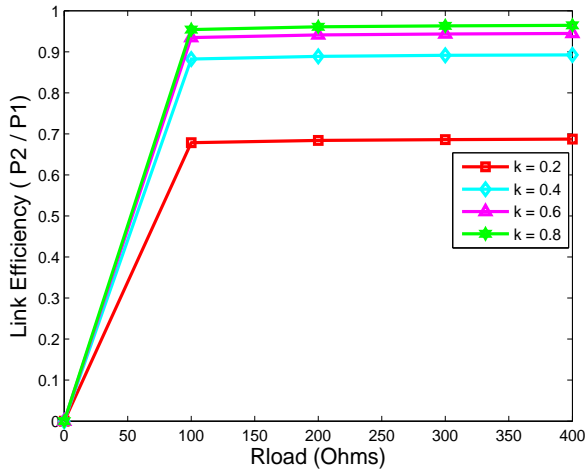}\\
  \caption{Link efficiency of series tuned primary and parallel tuned secondary circuit}\label{10}
\end{figure}

\section{Conclusion}
 Pacemakers are helpful in generating correct rhythms according to normal heart beats, in case of arrythmia. Whenever, the condition of Arrythmia occurs, it stores the information in it's hard drive, regarding time and event, thereby, helping the doctor to retrieve the information from the stored memory. In this study, pacemaker is used as an implanted WBAN device. Upon sensing any ambiguity in heart's functioning, it immediately routes the information towards the medical center. The transmission energy required, is more than the energy needed for driving the pacemaker. As a result, the lifetime of the battery of a sensor inside a pacemaker gets depleted earlier. To handle this issue, an induction technique is presented and analyzed, which re-charges the battery from time to time. Simulation results show that, the series tuned primary and parallel tuned secondary circuit achieves considerable good results for both voltage gain and link efficiency, in relation to the series tuned primary circuit.

\section{Future Work}
Authors in \cite{ain2012modeling} and \cite{sagar2012analysis}, derive analytical channel modeling and propagation characteristics of arm motion as spherical model. Future work includes a proposition of a mathematical induction model to recharge the sensors through an electric field which, is produced due to the arm movement in human body. These sensors will detect the arm motion. The focus is also to extend the work done in \cite{akbar2013modeling}, [15] and \cite{javaid2013srp} by proposing some techniques to recharge the sensors via induction. The induction will be achieved by considering mobility in the entire human body.

In \cite{hayat2012energy}, \cite{rahim2012adaptive}, \cite{rahim2012a} and \cite{alvi2012evaluation} authors discuss about MAC layer for BAN. In view of this, apart from achieving induction via mobility, future work also includes to work on cross layer designing and to propose induction techniques in such layer.

\ifCLASSOPTIONcaptionsoff
  \newpage
\fi

\end{document}